\documentclass[twocolumn,RiP,showpacs,superscriptaddress]{revtex4}%
\usepackage{amsfonts}
\usepackage{amsmath}
\usepackage{amssymb}
\usepackage{graphicx}%
\begin{document}
\title{Is equivalence principle valid for quantum gravitational field? }
\author{Baocheng Zhang}
\email{zhangbaocheng@cug.edu.cn}
\affiliation{School of Mathematics and Physics, China University of Geosciences, Wuhan
430074, China}

\begin{abstract}
Entanglement can be generated through the gravitational interaction between
two massive bodies that are initially in a product state. This shows that the
gravitational field is quantum. When the third massive body is introduced and
the gravitational interaction only between the third body with either one of
the former two bodies is considered, we find that no entanglement is generated
between the former two bodies up to the monopole approximation, even though
the considered gravitational interaction is quantum. This resembles the
behavior of two accelerating two-level atoms that is usually regarded as the
Unruh-DeWitt detectors. By linking the acceleration to that generated by the
gravitational field, we show that the equivalence principle is still valid
even though the gravitational field is quantum.

\end{abstract}
\maketitle

\section{Introduction}

Einstein equivalence principle (EEP) states that \textit{all effects of a
uniform gravitational field are identical to the effects of a uniformly
acceleration of a coordinate system }\cite{mtw73}. Thus, one cannot
distinguish a gravitational field from a uniformly accelerating frame of
reference by making any measurements in a small enough region of spacetime.
EEP is at the heart of the theory of General Relativity. Initially, Einstein
called it the \textquotedblleft equivalence hypothesis\textquotedblright\ when
he fully comprehended the importance of equivalence between gravitation and
inertia. In 1908, he elevated its name to the principle level as the
\textquotedblleft equivalence principle\textquotedblright\ after he realized
that this \textquotedblleft equivalence hypothesis\textquotedblright\ could be
used as a heuristic tool to build a physically satisfactory relativistic
theory of gravitation \cite{ae08,ae11}. In the past 100 years, there have been
many experiments to test the validity of EEP through the table-based or
space-based implementation \cite{cmw}. No evidence of violation are found
until now, but interest in such experiments still remains high due to its significance.

The EEP was constructed based on the classical theory in which the considered
elements were the classical massive objects and the classical gravitational
field. In recent years, however, some investigations \cite{zb18,ah18}
attempted to involve the quantum objects that may have internal degrees of
freedom instead of the classical objects to discuss whether EEP is still valid
in such a case. These studies aimed to construct a general equivalence
principle that is also proper for quantum objects, by which it is expected to
search for the possibility or the clues of constructing the theory of quantum
gravity. Some experiments using neutron \cite{co75,bw83} or atomic
interferometry \cite{fdw04,hdl12} has tested such ideas, and no violation is
found until now. Nevertheless, it is hardly discussed whether the EEP is valid
when the classical gravitational field is substituted by the quantum
gravitational field. The reason is that no one knows how to describe the
quantum gravitational field properly due to the absence of the complete theory
about quantum gravity. However, there exists some methods to test gravity's
quantum behavior but don't require any particular theory of quantum gravity in
advance, as pointed out in recent works \cite{mv17,bmt17}. Based on this, here
we will study the question whether the EEP is valid for the quantum
gravitational field.

\section{Gravitational field}

According to the thought from Ref. \cite{mv17,bmt17}\ and as stated in Ref.
\cite{mvn17}, \textit{if gravity follows quantum theory, it should set into a
superposition of many states at once when it interacts with a mass that is
also behaving in this way. A second mass could be used as a probe to pick up
that quantum state}. \textit{Measuring the probe's state could determine
whether it has been superposed, thus proving whether gravity exhibits quantum
behavior.} This proposed theoretically the methods for the experimental
confirmation of the assumption that the gravitational field is quantum by the
result that either it is in a superposed state or it generates entanglement
between two massive bodies. We will continue along this line. For simplicity,
two quantum states are considered to be complete for the description of the
gravitational field in this paper. At first, we provide a way that has the
same evolution essentially as before \cite{mv17,bmt17}, that is,
\begin{equation}
\left\vert m_{1}\right\rangle \left\vert m_{1}^{\prime}\right\rangle
\rightarrow\alpha_{1}\left\vert m_{1}\right\rangle \left\vert m_{1}^{\prime
}\right\rangle +\sqrt{P}\alpha_{2}\left\vert m_{2}\right\rangle \left\vert
m_{2}^{\prime}\right\rangle , \label{fef}%
\end{equation}
where two massive bodies separated by a distance $d$ are distinguished by the
primed and unprimed labels. As stated in Ref. \cite{mvn17}, the quantum
behavior of the gravitational field can be described by analogy from the
massive body that produced it. Thus, the quantum states $\left\vert
m_{1}\right\rangle $ and $\left\vert m_{2}\right\rangle $ for one massive body
(or for the gravitational field it produces) can be regarded as energy
eigenstates like that usually used in the quantum theory, and $m_{1}$ or
$m_{2}$ is the mass-like parameter related to the energy of the corresponding
state by the Einstein mass-energy relation \cite{mtw73}. The primed labels
have the same meaning for another massive body. $\alpha_{1}\sim\exp
[-i\frac{Gm_{1}m_{1}^{\prime}}{d}\frac{t}{\hbar}]$ and $\alpha_{2}\sim
\exp[-i\frac{Gm_{2}m_{2}^{\prime}}{d}\frac{t}{\hbar}]$ represent the phases
derived from the gravitational interaction.\ $P$ represents the probability
for the change of the state from $\left\vert m_{1}\right\rangle \left\vert
m_{1}^{\prime}\right\rangle $ to $\left\vert m_{2}\right\rangle \left\vert
m_{2}^{\prime}\right\rangle $ through the gravitational interaction between
them. Evidently, the initial product state becomes entangled after the
interaction, which shows that the gravitational interaction is quantum. This
is a simple presentation for the earlier proposals
\cite{mv17,bmt17,mvn17,mvd18} in which the interaction is exerted at such a
product state $\left(  \frac{\left\vert m_{1}\right\rangle +\left\vert
m_{2}\right\rangle }{\sqrt{2}}\right)  \left(  \frac{\left\vert m_{1}^{\prime
}\right\rangle +\left\vert m_{2}^{\prime}\right\rangle }{\sqrt{2}}\right)  $.
It is easy to recover the earlier results with the evolution presented in Eq.
(\ref{fef}) up to a normalization factor.

Now we consider a gravitational field $G$ generated by the Earth. It can be
described by two quantum states $\left\vert M_{1}\right\rangle $ and
$\left\vert M_{2}\right\rangle $ simply as stated above. An atom is placed
into the field $G$ at height $h$ to \textquotedblleft feel\textquotedblright%
\ the interaction. The atom is regarded as a point-like two-level quantum
system with the ground $\left\vert g\right\rangle $ and excited $\left\vert
e\right\rangle $ states. According to Eq. (\ref{fef}), the coupling between
the atom and the field $G$ is expressed as%
\begin{align}
U^{G}\left\vert g\right\rangle \left\vert M_{1}\right\rangle  &  =C_{0}\left(
\alpha_{1}^{G}\left\vert g\right\rangle \left\vert M_{1}\right\rangle
+\sqrt{P_{ge}^{G}}\alpha_{2}^{G}\left\vert e\right\rangle \left\vert
M_{2}\right\rangle \right)  ,\nonumber\\
U^{G}\left\vert e\right\rangle \left\vert M_{1}\right\rangle  &  =C_{1}\left(
\beta_{1}^{G}\left\vert e\right\rangle \left\vert M_{1}\right\rangle
+\sqrt{P_{eg}^{G}}\beta_{2}^{G}\left\vert g\right\rangle \left\vert
M_{2}\right\rangle \right)  , \label{eai}%
\end{align}
where $C_{0,1}$ is the state normalization factor, and $U^{G}$ represents the
interaction between the atom and the gravitational field. The interaction
derives from the quantization of the Hamiltonian $H_{int}^{G}=-\frac{1}%
{2}h_{\mu\nu}T^{\mu\nu\text{ }}$where $T^{\mu\nu\text{ }}$is the stress-energy
tensor and $h_{\mu\nu}$ is the perturbation of the metric tensor away from the
flat (Minkowski) spacetime. In the earlier proposals \cite{mv17,bmt17}, the
quantum states of gravitational field were described as spatially localized
states, i.e. the path states separated by up and down in Ref. \cite{mv17}\ and
that separated by left and right in Ref. \cite{bmt17}. Here the states
$\left\vert M_{1}\right\rangle $ and $\left\vert M_{2}\right\rangle $ are
essentially the same with the earlier usage, but it might be better to be
understood roughly as the energy eigenstates of the gravitational field. For
example, the atom absorbs the energy from the field $G$, which would lead to
the change of the energy or the state of the field $G$. Because the mass of
the Earth is much larger than the atom, the quantum states corresponding to
its gravitational field are dense in this sense that $\frac{\left\vert
M_{2}-M_{1}\right\vert }{M_{1}}\ll\frac{\left\vert m_{e}-m_{g}\right\vert
}{m_{g}}$. Moreover, the states $\left\vert M_{1}\right\rangle $ and
$\left\vert M_{2}\right\rangle $ can be orthogonal only if the quantum
gravitational field has a complete set of state vectors. As stated above, we
consider here that the number of state vectors for the set is two for
simplify. The parameters $\alpha_{1}^{G}$, $\alpha_{2}^{G}$, $\beta_{1}^{G}$,
and $\beta_{2}^{G}$ can be obtained according to the Eq. (\ref{fef}).
$P_{ge}^{G}$ ($P_{eg}^{G}$) represents the transition probability from ground
(exited) to exited (ground) state for the atom interacting with the field $G$.

When two atoms $A$ and $B$ with the initial product state $\left\vert
g\right\rangle \left\vert e\right\rangle $ are placed into the gravitational
field $\left\vert M_{1}\right\rangle $ at the same height but different sites,
through the interaction with the gravitational field $G$, the state becomes%
\begin{align}
|\Psi_{f}^{G}\rangle &  =\left(  U^{G}\left\vert g\right\rangle \left\vert
M_{1}\right\rangle \right)  \left(  U^{G}\left\vert e\right\rangle \left\vert
M_{1}\right\rangle \right) \nonumber\\
&  =C_{0}C_{1}[\alpha_{1}^{G}\beta_{1}^{G}\left\vert g\right\rangle \left\vert
e\right\rangle \left\vert M_{1}\right\rangle \left\vert M_{1}\right\rangle
\nonumber\\
&  +\sqrt{P_{eg}^{G}}\alpha_{1}^{G}\beta_{2}^{G}\left\vert g\right\rangle
\left\vert g\right\rangle \left\vert M_{1}\right\rangle \left\vert
M_{2}\right\rangle \nonumber\\
&  +\sqrt{P_{ge}^{G}}\alpha_{2}^{G}\beta_{1}^{G}\left\vert e\right\rangle
\left\vert e\right\rangle \left\vert M_{1}\right\rangle \left\vert
M_{2}\right\rangle \nonumber\\
&  +\sqrt{P_{ge}^{G}}\sqrt{P_{eg}^{G}}\alpha_{2}^{G}\beta_{2}^{G}\left\vert
e\right\rangle \left\vert g\right\rangle \left\vert M_{1}\right\rangle
\left\vert M_{2}\right\rangle ] \label{ces}%
\end{align}
Tracing out the quantum states of the gravitational field, we obtain the final
state for the two atoms as%
\begin{equation}
\rho_{f}^{G}=\rho_{A}^{G}\otimes\rho_{B}^{G}, \label{fir}%
\end{equation}
where $\rho_{A}=$ $\left\vert C_{0}\right\vert ^{2}\left(  \left\vert
\alpha_{1}^{G}\right\vert ^{2}\left\vert g\right\rangle \left\langle
g\right\vert +P_{ge}^{G}\left\vert \alpha_{2}^{G}\right\vert ^{2}\left\vert
e\right\rangle \left\langle e\right\vert \right)  $ and $\rho_{B}=$
$\left\vert C_{1}\right\vert ^{2}\left(  \left\vert \beta_{1}^{G}\right\vert
^{2}\left\vert e\right\rangle \left\langle e\right\vert +P_{eg}^{G}\left\vert
\beta_{2}^{G}\right\vert ^{2}\left\vert g\right\rangle \left\langle
g\right\vert \right)  $. It shows that no entanglement is created when the two
atoms are coupled to the same gravitational field. This is obtained by
ignoring the gravitational interaction between two atoms, since the
interaction is much smaller than that between the atom and the Earth. But in
the earlier proposals \cite{mv17,bmt17}, the generation of entanglement is due
to the gravitational interaction between two atoms. This is embodied here in
the generation of entanglement between the atom and the Earth, which
guarantees that the gravitational field $G$ is quantum. Moreover, when the two
atoms are placed in the different height, the similar result can be obtained.

\section{Acceleration}

In order to investigate the EEP, we turn to the effects in an accelerating
frame of reference. It is widely accepted by now that an observer with the
uniform acceleration $a$ in the Minkowski vacuum would feel a thermal bath of
particles at the temperature $T_{U}=\hbar a/2\pi ck_{B}$. This is just the
well-known Unruh effect \cite{wgu76}. It has been digested and extended to
many different situations (see the review \cite{chm08} and references
therein). In one famous application to the Unruh-DeWitt (UDW) detector
\cite{dhi79}, it is found that a quantum system consisting of a detector
uniformly accelerating in Minkowski vacuum can sense the thermal emission and,
thus, cause the coupling of the quantum system with the thermal field. In what
follow, we will make the discussion using the UDW model.

Starting with the consideration that constitutes a scalar field $\phi$
interacting with a point-like two-level quantum system, or an atom as used in
the discussion above. The interaction Hamiltonian for this ($1+1$)-dimension
model can be modeled by the interaction Hamiltonian, $H_{I}=\lambda\mu\left(
\tau\right)  \phi\left(  x\left(  \tau\right)  \right)  $, with $\lambda$ the
coupling strength. $\tau$ is the atom's proper time along its trajectory
$x\left(  \tau\right)  $, $\mu\left(  \tau\right)  $ is the atom's monopole
momentum, and $\phi(x(\tau))$ is the scalar field related to the vacuum. For
an atom accelerating in the Minkowski vacuum, the evolution of the total
quantum state is determined perturbatively by the unitary operator which up to
first order is given by, $U=I+U^{(1)}+O\left(  \lambda^{2}\right)  =I-i\int
d\tau H\left(  \tau\right)  +O\left(  \lambda^{2}\right)  $. Within the
first-order approximation and in the interaction picture, this evolution is
described by \cite{bmm16,lzy18}
\begin{align}
U\left\vert g\right\rangle \left\vert 0\right\rangle  &  =D_{0}\left(
\left\vert g\right\rangle \left\vert 0\right\rangle +\eta_{_{0}}\left\vert
e\right\rangle \left\vert 1_{k}\right\rangle \right)  ,\nonumber\\
U\left\vert e\right\rangle \left\vert 0\right\rangle  &  =D_{1}\left(
\left\vert e\right\rangle \left\vert 0\right\rangle +\eta_{_{1}}\left\vert
g\right\rangle \left\vert 1_{k}\right\rangle \right)  , \label{foe}%
\end{align}
where $k$ denotes the mode of the ($1+1$)-dimension scalar field with
(bosonic) annihilation (creation) operator $a_{k}$ ($a_{k}^{\dag}$),
$a_{k}\left\vert 0\right\rangle =0$ and $a_{k}^{\dag}\left\vert 0\right\rangle
=\left\vert 1_{k}\right\rangle $. $D_{0,1}$ is the state normalization factor.
$\eta_{_{0}}$ and $\eta_{_{1}}$ are related to the excitation and deexcitation
probability of the atom, i.e. $P_{ge}=$ $\sum_{k}\left\vert \left\langle
e,1_{k}\right\vert U^{(1)}\left\vert g,0\right\rangle \right\vert
^{2}=\left\vert \eta_{_{0}}\right\vert ^{2}$ and $P_{eg}=$ $\sum_{k}\left\vert
\left\langle g,1_{k}\right\vert U^{(1)}\left\vert e,0\right\rangle \right\vert
^{2}=\left\vert \eta_{_{1}}\right\vert ^{2}$ (the relative phases are absorbed
into the parameters $\eta_{_{0}}$ and $\eta_{_{1}}$ whose concrete form refers
to Ref. \cite{chm08,bmm16}). $t\left(  \tau\right)  =\frac{c}{a}\sinh
(\frac{a\tau}{c})$ and $x\left(  \tau\right)  =\frac{c^{2}}{a}\left(
\cosh(\frac{a\tau}{c})-1\right)  $ is the trajectory of the accelerating atom
with acceleration $a$. Approximately, $t\simeq\tau$ and $x\simeq\frac{1}%
{2}a\tau^{2}$ indicates the trajectory of uniformly accelerated rectilinear motion.

When two atoms $A$ and $B$ with the initial product state $\left\vert
g\right\rangle \left\vert e\right\rangle $ are accelerated with the same
acceleration, and the initial product state then evolves into%
\begin{align}
&  |\Psi_{f}\rangle=D_{0}D_{1}(\left\vert g\right\rangle \left\vert
e\right\rangle \left\vert 0\right\rangle \left\vert 0\right\rangle +\eta
_{_{1}}\left\vert g\right\rangle \left\vert g\right\rangle \left\vert
0\right\rangle \left\vert 1_{k}\right\rangle \nonumber\\
&  +\eta_{_{0}}\left\vert e\right\rangle \left\vert e\right\rangle \left\vert
1_{k}\right\rangle \left\vert 0\right\rangle +\eta_{_{0}}\eta_{_{1}}\left\vert
e\right\rangle \left\vert g\right\rangle \left\vert 1_{k}\right\rangle
\left\vert 1_{k}\right\rangle ).
\end{align}
Again, the reduced two-atom density matrix is obtained by tracing out the
scalar field modes
\begin{equation}
\rho_{f}=\rho_{A}\otimes\rho_{B}, \label{aas}%
\end{equation}
It shows that no entanglement is created. This is the same as the result
(\ref{fir}) obtained from the quantum gravitational field. Moreover, when the
two atoms assume different accelerations, the final quantum state is found to
take a similar form to that Eq. (\ref{aas}) for two atoms with the same acceleration.

\section{Equivalence between gravitation and acceleration}

Then, let's see how to relate the gravitational field with the acceleration.
As well-known, the accelerating frame of reference can be expressed with the
Rindler metric $ds^{2}=\rho^{2}d\xi^{2}-d\rho^{2}$ \cite{chm08}. It might be
easier to understand this by using the associated coordinates $z=\rho
-\frac{c^{2}}{a}$ and $\tau=\frac{\rho}{c}\xi$, in terms of which the Rindler
metric becomes \cite{gf87}%
\begin{equation}
ds^{2}=-\left(  1+\frac{az}{c^{2}}\right)  ^{2}c^{2}d\tau^{2}+dz^{2}%
,\label{rma}%
\end{equation}
where $\tau$ represents the proper time, and $a=\frac{c^{2}}{\rho}$ (at the
point $z=0$) represents the acceleration. This coordinates can also be
interpreted with the static gravitational field where $g=-a$ is the
acceleration of free fall at that point with respect to the chosen body at
rest. In the Newtonian approximation, when the potential $\phi\ll c^{2}$, one
has%
\begin{align}
ds^{2} &  =-\left(  1+\frac{2\phi}{c^{2}}\right)  ^{2}c^{2}d\tau^{2}+\left(
1-\frac{2\phi}{c^{2}}\right)  dz^{2}\nonumber\\
&  \simeq-\left(  1+\frac{\phi}{c^{2}}\right)  ^{2}c^{2}d\tau^{2}+dz^{2},
\end{align}
up to a conformal factor in the last step. From this, it can be seen that the
metric (\ref{rma}) corresponds to the case of a homogeneous gravitational
field (in the paper, we consider the atoms rested in the specific points of
the gravitational field, and thus for the small enough regime around the
points, the field can be regarded as homogeneous) in the Newtonian
approximation with the potential $\phi=-gz=az$ which provides the relation
between the description of accelerated atoms and the atoms rested in the
gravitational field.

For the accelerated atom, the transition ratio can be calculated as
$\frac{P_{ge}}{P_{eg}}\simeq$ $\exp[-\frac{\Delta E}{k_{B}T_{U}}]=\exp
[-\frac{2\pi c\Delta E}{a}]$ for a long enough time, where $\Delta E$ is the
energy gap between the ground $\left\vert g\right\rangle $ and excited
$\left\vert e\right\rangle $ states of the atom. It indicates the thermal
response of a particle detector evaluated by the detailed-balance condition
obeyed by KMS states \cite{st86}. Similarly, the transition ratio caused by
the gravitational field can be calculated as $\frac{P_{ge}^{G}}{P_{eg}^{G}%
}=\exp[\frac{2\pi c\Delta E}{g}]$, which derives from the consideration
\cite{gf87} that a detector at rest in a static homogeneous gravitational
field at point at which the acceleration of free fall is $g=-a$ behaves in the
case of interaction with the quantized field in the Hartle-Hawking vacuum
state \cite{hh76}. The Hartle-Hawking vacuum state is related to the thermal
equilibrium state in the free falling frame of reference in the static
gravitational field, while the Boulware vacuum \cite{dgb75} is seen for
observers rested in the static gravitational field. This confirms that the EEP
is still valid when the gravitational field is regarded as quantum, although
the calculation is made in the monopole approximation both for the gravity and
the acceleration. When the higher expansion is considered, entanglement could
be generated according to the process stated above, which is similar to the
mechanism of entanglement harvesting \cite{smm15} in the acceleration. 

Notably, the connection between acceleration and gravity relies on the
transition probability of a single atom. However, our validation of the EEP is
grounded in the entanglement of two atoms. This extension holds true as,
within the considered approximation, verifying the EEP necessitates that the
two atoms experience identical gravitational acceleration and are subjected to
the same acceleration in a flat spacetime background. The entanglement under
discussion is intricately associated with the transition probability of each
atom, allowing the confirmation of the EEP through the alteration of
entanglement in distinct scenarios.

\section{Conclusion}

In conclusion, the EEP is discussed under the background of quantum
gravitational field. The quantum property of the gravitational field is
confirmed by generating entanglement between the source (Earth) and the
objects (atoms). When two atoms are placed in the gravitational field and
interact with the field respectively, we show that no entanglement is
generated between the two atoms. This conclusion is obtained by taking the
interaction in the monopole approximation and ignoring the interaction between
two atoms since the mass of the atom is much smaller than that of the Earth.
On the other hand, when two atoms are accelerated in the Minkowski vacuum, no
entanglement is generated either under the interaction of the first-order
approximation. Thus, by linking the acceleration to the static homogeneous (at
least locally) gravitational field, we show that the EEP is still valid for
the quantum gravitational field up to the monopole approximation. More general
confirmation for the EEP might require more knowledge about the vacuum and the
high-order interaction for the strong gravitational field \cite{sw11}, and
more refined quantum criteria for the gravitational field \cite{hr18}.

\section{Acknowledgements}

This work is supported by National Natural Science Foundation of China (NSFC)
with Grant No. 12375057, and the Fundamental Research Funds for the Central
Universities, China University of Geosciences (Wuhan) with No. G1323523064.


\begin{thebibliography}{99}                                                                                               %


\bibitem {mtw73}C. W. Misner, K. S. Thorne, and J. A. Wheeler,
\textit{Gravitation} (W. H. Freeman and Company, San Francisco, USA, 1973).

\bibitem {ae08}A. Einstein, Jahrbuch Radiaoktiv. 4, 411 (1908).

\bibitem {ae11}A. Einstein, Ann. Phys. Lpz., 35, 898 (1911).

\bibitem {cmw}C. M. Will, Living Rev. Relativity, 9, 3 (2006).

\bibitem {zb18}M. Zych and \v{C}. Brukner, Nature Phys 14, 1027--1031 (2018).

\bibitem {ah18}C. Anastopoulos and B. L. Hu, Class. Quantum Grav. 35, 035011 (2018).

\bibitem {co75}R. Colella, A. Overhauser, and S. Werner, Phys. Rev. Lett. 34,
1472--1474 (1975).

\bibitem {bw83}U. Bonse and T. Wroblewski, Phys. Rev. Lett. 51, 1401 (1983).

\bibitem {fdw04}S. Fray, C. A. Diez, T. W. H\"{a}nsch, and M. Weitz, Phys.
Rev. Lett. 93, 240404 (2004).

\bibitem {hdl12}S. Herrmann, H. Dittus, C. L\"{a}merzahl, et al., Classical
and Quantum Gravity 29, 184003 (2012).

\bibitem {mv17}C. Marletto and V. Vedral, Phys. Rev. Lett. 119, 240402 (2017).

\bibitem {bmt17}Sougato Bose, Anupam Mazumdar, Gavin W. Morley,\ et al., Phys.
Rev. Lett. 119, 240401 (2017).

\bibitem {mvn17}C. Marletto and V. Vedral, Nature 547, 156 (2017).

\bibitem {mvd18}C. Marletto, V. Vedral, and D. Deutsch, New J. Phys. 20,
083011 (2018).

\bibitem {wgu76}W. G. Unruh, Phys. Rev. D 14, 870 (1976).

\bibitem {chm08}L. C. B. Crispino, A. Higuchi, and G. E. A. Matsas, Rev. Mod.
Phys. 80, 787 (2008).

\bibitem {dhi79}B. S. DeWitt, S. Hawking, and W. Israel, \textit{General
Relativity: An Einstein Centenary Survey} (Cambridge University Press
Cambridge, England, 1979).

\bibitem {bmm16}W. Brenna, R. B. Mann, and E. Mart\'{\i}n-Mart\'{\i}nez, Phys.
Lett. B 757, 307 (2016).

\bibitem {lzy18}T. Li, B. Zhang and L. You, Phys. Rev. D 97, 045005 (2018).

\bibitem {gf87}V. L. Ginzburg and V. P. Frolov, Sov. Phys. Usp. 30 1073 (1987).

\bibitem {st86}S. Takagi, Prog. Theor. Phys. Suppl. 88, 1 (1986).

\bibitem {hh76}J. B. Hartle and S. W. Hawking, Phys. Rev. D 13, 2188 (1976).

\bibitem {dgb75}D. G. Boulware, Phys. Rev. D 11, 1404 (1975).

\bibitem {smm15}G. Salton, R. B. Mann, and N. C. Menicucci, New J. Phys. 17,
035001 (2015).

\bibitem {sw11}D. Singleton and S. Wilburn, Phys. Rev. Lett. 107, 081102 (2011).

\bibitem {hr18}M. J W Hall and M. Reginatto, J. Phys. A: Math. Theor. 51,
085303 (2018).
\end{thebibliography}
\end{document}